\documentclass[aps,twocolumn,showpacs]{revtex4}
\usepackage{graphics}

\begin{document}

\title{\Large{Effect of external electric field on the charge density waves in
    one dimensional Hubbard superlattices}}
\author{Jayeeta Chowdhury$^1$}
\author{S. N. Karmakar$^2$}
\author{Bibhas Bhattacharyya$^3$}
{\affiliation{$^1$Department of Physics, East Calcutta Girls' College,\\
 P-237, Block B, Lake Town, Lake Town Link Road, Kolkata 700 089, India\\
$^2$TCMP Division, Saha Institute of Nuclear Physics, 1/AF Bidhannagar, Kolkata 700 064, India\\
$^3$Department of Physics, Jadavpur University, Kolkata 700 032, India}

\begin{abstract}
We have studied the ground state of the one dimensional Hubbard superlattice
structures with different unit cell sizes in the presence of electric
field. Self consistent Hartree-Fock approximation calculation is done in the
weak to intermediate interaction regime. Studying the charge gap at the Fermi
level and the charge density structure factor, we get an idea how the charge
modulation on the superlattice is governed by the competition between the
electronic correlation and the external electric field.
\end{abstract}

\pacs{73.21.Cd, 71.30.+h, 71.45.Lr, 74.25.Jb}

\maketitle

\section{Introduction}
The study of low dimensional metallic multilayered structures \cite{Heinrich}
is interesting because of their unique characteristics. The oscillation of
exchange coupling between magnetic layers \cite{Grunberg} and the appearance
of giant magnetoresistance \cite{Baibich} are among the exciting features of
the multilayers. To investigate the properties of the metallic multilayers
many theoretical works have been done taking simple superlattice structures as
the models \cite{Santos1,Santos3,Santos4,Chow,Santos2}. This kind of models
consist of periodic arrangement of $N_U$ sites with repulsive on-site
Coulomb interaction $U$ $(>0)$, followed by $N_0$ sites with no on-site
interaction $(U=0)$. Some of these works investigated the ordering 
\cite{Santos1,Santos3,Santos4,Chow} of the ground state, while some other
explored the
possibility of the metal-insulator transition in these systems \cite{Santos2,
  Chow}. There are possibilities of formation of novel ground states such as
charge ordered or spin ordered ones depending on the distribution of the
interaction parameter $U$ in such superlattices.\\

On the other hand, the effect of electric field on the strongly correlated low
dimensional  electronic systems has attracted much interest in recent years
because of their practical applications in tuning dielectric and piezoelectric
properties \cite{Pati1}. Many experiments are done on these low dimensional
systems in the presence of  electric field. It was found that spin ordered or
charge  ordered phases of a Mott insulator collapse in an electric
field \cite{Taguchi,Asamitsu,Rao,Wu,Dumas}. Also some theoretical works are 
done on such systems in which a uniform electric field is implemented in the 
form of a ramp potential. Applying such an electric field in the homogeneous
Hubbard model it was found that the field can induce oscillations in the
charge gap of these systems \cite{Pati2,Pati3}. However, it is not yet known
how the superlattice systems behave in the presence of such an electric 
field.\\

In this work, we investigate the electronic properties of simple superlattice
structures in the presence of electric field. We consider weak to intermediate 
interaction regime and work under the Hartree-Fock Approximation (HFA). \\

\section{The model and the Hartree-Fock Approximation}

Our model is a  one dimensional $N$-site Hubbard chain with open
boundaries. The model Hamiltonian is,
\begin{equation}
H=\sum_{i}\epsilon_{i}n_{i}+t\sum_{i,\sigma}(c_{i,\sigma}^{\dag}c_{i+1,\sigma}+H.c.)+\sum_{i}U_{i}n_{i,\uparrow}n_{i,\downarrow},~~
\end{equation}

\noindent 
where $c_{i,\sigma}^{\dag}(c_{i,\sigma})$ is the creation(annihilation)
operator for an electron with spin $\sigma$ ($\uparrow$ or $\downarrow$) at
the $i$-th site. $n_{i,\sigma}=c_{i,\sigma}^{\dag}c_{i,\sigma}$, and
$n_{i}=\sum_{\sigma}n_{i,\sigma}$ is the number operator at the $i$-th site;
$t$ is the hopping integral between the nearest neighbor sites.
$U_{i}$ denotes the on-site Coulomb repulsion at the $i$-th site; in a 
superlattice $U_{i}$'s follow a repeated pattern
depending on the size of a unit cell of the superlattice. $\epsilon_{i}$ is the
site energy of the $i$-th site. In the absence of electric field all   
$\epsilon_{i}$'s are set to zero. The external electric field is applied on
the system in the form of a ramp potential \cite{Pati2}. In presence of this
field the site energies become

\[\epsilon_{i}=-\frac{V}{2}+i \frac{V}{N+1},\]

\noindent
where $V$ is the applied
voltage. This form of the site potential is used to ensure that the external 
bias varies from $-V/2$ to $V/2$ across the superlattice. We will work in the
 weak to intermediate coupling regime where
$U {< \atop \sim} t$. It was observed in a previous work that in this
regime the mean field approximation is quite reliable for this class of  
systems \cite{Chow}.\\

We decouple the Hamiltonian using the unrestricted Hartree-Fock Approximation
(HFA), 
\begin{equation}
U n_{i,\uparrow} n_{i,\downarrow} \rightarrow U \langle n_{i,\uparrow}\rangle
n_{i,\downarrow}+U n_{i,\uparrow} \langle n_{i,\downarrow}\rangle -U \langle
n_{i,\uparrow} \rangle \langle n_{i,\downarrow} \rangle,
\end{equation}

\noindent
where $\langle \cdots \rangle$ denotes the expectation value calculated with 
respect to the ground state. Now the Hamiltonian can be divided into two parts 
for the two types of spins, i.e. ~$H=H_{\uparrow}+H_{\downarrow}$. In an
unrestricted Hartree-Fock Approximation, one determines the distribution of
the $n_{i,\sigma}$'s by diagonalizing $H_{\uparrow}$ and $H_{\downarrow}$ in a 
self-consistent manner. The ground state is constructed by filling up the
energy levels from both the up and the down bands upto the Fermi level.\\

\begin{figure}[h]
\resizebox{8cm}{!}
{\includegraphics*{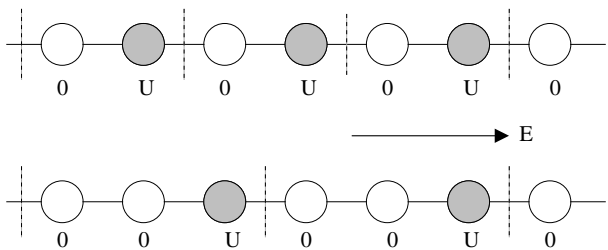}}
\caption{Two different types of superlattice structures studied in the present
  work. The arrow shows the direction of the electric field.}
\end{figure}

In this paper, we have presented the results for two different types of
superlattice structures as shown in Fig.~$1$. For the first one the size of
the unit cell is two and for the second one it is three. We have taken two
types of sites with on-site correlation parameters $U=1$ and $U=0$
respectively. We have studied the superlattices for various  values of
$N$. Since our aim is to compare the effects of the electric field on
different superlattice structures of small size, we have presented here the
results for the cases with $N=60$ and $120$ only. Comparing the results for 
these two system sizes one can also make an idea what happens in the infinite 
limit.\\ 

To study the effect of
the electric field on the metallic/insulating behavior of the ground state, we 
performed a systematic study of the
charge gap $(\Delta)$ at the Fermi level of the system containing $n$ 
electrons,
\begin{equation}
\Delta=E_{n+1}+E_{n-1}-2E_{n},
\end{equation}
where $E_{n}$ is the ground state energy of an $n$ electron system. We have
also studied the charge density wave (CDW) structure factor
\begin{equation}
C(q)=\frac{1}{N}\sum_{i,j} e^{iq(r_{i}-r_{j})}(n_{i}-\rho)(n_{j}-\rho),
\end{equation}
 where $\rho$ is the average particle density on the superlattice,
 $r_p$ denotes the position of the $p$-th site and $q$ is the wave
 vector. These two quantities enable us to capture the competition between the
 electric field and the correlation parameter in determining the charge
 modulation along the superlattice.\\

\section{Results of HFA calculations}

It is well known that in the absence of electric field, a homogeneous Hubbard 
chain is an anti-ferromagnetically ordered system with a finite charge gap at 
half-filling \cite{Lieb,Shiba}. Figure~$2$ shows the 
variation of the charge 
gap with the electric field for a half-filled homogeneous Hubbard
chain. It is clear from the diagram that the charge gap goes
through a number of maxima and minima with increasing electric field. As $N$ 
increases the minima shift towards lower values of $V$. The
nature of variation of the charge gap is in good qualitative agreement with
the previous Density Matrix Renormalization Group (DMRG) results \cite{Pati2}.\\

\begin{figure}[h]
\resizebox{8cm}{!}
{\includegraphics*{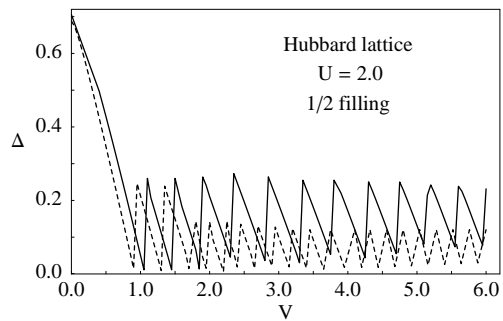}}
\caption{Variation of the charge gap $(\Delta)$ at the Fermi level for the
  homogeneous Hubbard chain at half-filling with the applied voltage $(V)$. 
The solid line corresponds to $N=60$, while the dotted line shows the case of 
$N=120$.}
\end{figure}

\begin{figure}[h]
\resizebox{7.5cm}{!}
{\includegraphics*{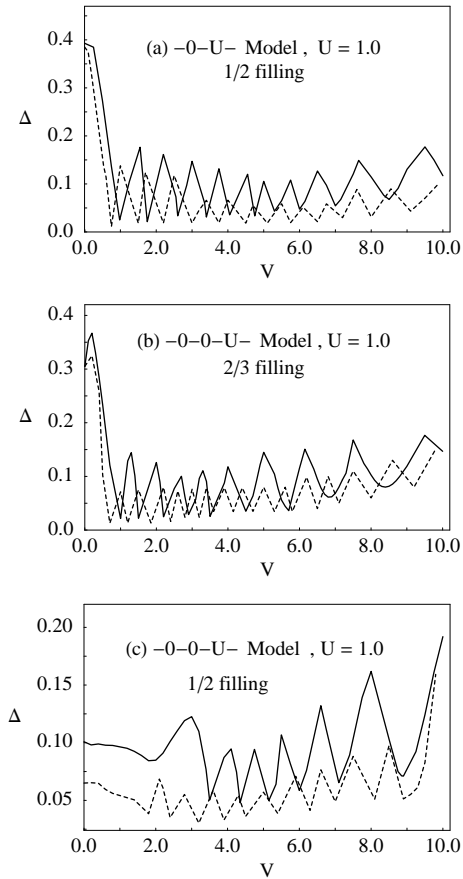}}
\caption{Variations of the charge gap $(\Delta)$ at the Fermi level of the
  superlattice systems with the applied voltage $(V)$; (a) for $-0-U-$
  superlattice at $\frac{1}{2}$ filling, (b) for $-0-0-U-$ superlattice at
  $\frac{2}{3}$ filling and (c) for  $-0-0-U-$ superlattice at
  $\frac{1}{2}$ filling respectively. The solid line corresponds to $N=60$,
 while the dotted line shows the case of 
$N=120$.}
\end{figure}

Now we discuss the results for the $-0-U-$ superlattice structure. At
half-filling the system is a CDW insulator with $q=\pi$ in the absence of
electric field \cite{Santos3, Chow} and the charge gap is finite
($V=0$ case in Fig.~$3(a)$). As we turn on and gradually increase the
electric field, the charge gap passes through a number
of maxima and minima. For $N=60$ and $120$the variation of the charge gap 
are shown in Fig.~$3(a)$. The type of oscillation of the charge
gap with electric field is quite similar to that observed in the homogeneous
system (Fig.~$2$). These oscillations are  observable only in the finite sized
systems and are crucially  controlled by the interplay of the Hubbard
interaction and the spatial gradient of the external bias. In the absence of 
electric field, the CDW phase (with $q=\pi$) at
half-filling has tendency to form ``doublons'' $(\uparrow \downarrow)$ at the
sites with $U=0$, while the sites with $U>0$ tend to depopulate. When the
electric field is increased the electrons are pushed back near one end of the
lattice. This leads to population of some sites with $U>0$, leading to
the breakdown of the $q=\pi$ CDW phase. Such a crossover from the CDW
phase is marked by the first minimum in the charge gap. Subsequent increase in
the electric field results in gradual accumulation of electrons in one half of
the superlattice. Because of the competition between the Coulomb correlation
energy and the electric field term, the aforesaid process of piling up of
electrons takes place only after finite increments of electric field in a
finite sized system. This results in the oscillation in the charge gap.\\

\begin{figure}[h]
\resizebox{7.5cm}{!}
{\includegraphics*{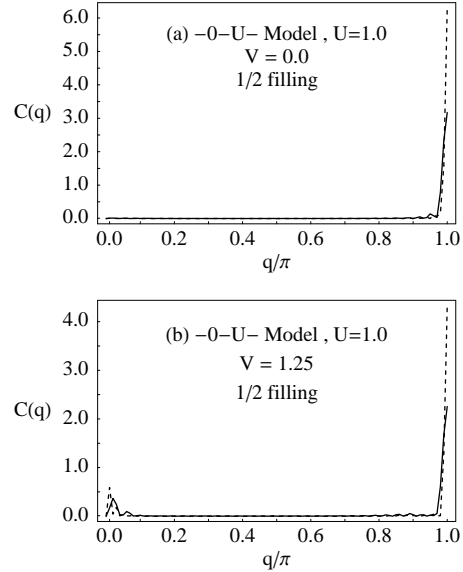}}
\caption{Charge density structure factor $C(q)$ for the half filled $-0-U-$
  superlattice systems; (a) for $V=0.0$ and (b) for $V=1.25$ respectively.
 The solid line corresponds to $N=60$, while the dotted line represents the 
case of $N=120$.}
\end{figure}

\begin{figure}[h]
\resizebox{7.5cm}{!}
{\includegraphics*{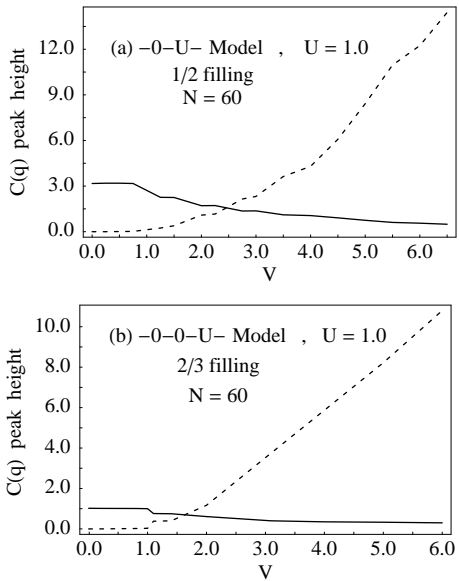}}
\caption{Variations of the height of the peaks in $C(q)$ for the
  superlattice systems with the applied voltage $(V)$; dashed line is for
  the peak at $q\simeq 0$, while solid line is for the peak at $q \simeq \pi$
  and $q \simeq \frac{2\pi}{3}$ in (a) $-0-U-$ superlattice at
  $\frac{1}{2}$ filling and (b) $-0-0-U-$ superlattice at $\frac{2}{3}$
  filling respectively.} 
\end{figure}

Next we study the CDW structure factor of this model. Generally speaking, from
the mean field point of view, all sites with the same value of $U$ in a
superlattice are of same status. So a peak depending on the periodicity of the
structure of the superlattice appears in the charge density structure
factor. Our method also detects other peaks in $C(q)$'s that depend on the
specific values of the density (i.e. the position of the peak depends on the
Fermi wave vector $k_F$). As the system size $N$ increases, the peak due to
the structural periodicity of the lattice becomes larger compared to the other
peaks. Under periodic boundary condition or in the limit of $N \rightarrow
\infty$, only the peak due to the structural modulation survives and the 
other peaks disappear.\\

Referring to the specific case of  $-0-U-$
superlattice at half-filling, there is a sharp peak in
$C(q)$ at $q \simeq \pi$ in the absence of electric field
(Fig.~$4(a)$). Incidentally, in this case the peak due to the periodicity of
the lattice structure and the $2k_{F}$ peak both occur at $q \simeq \pi$. As
the electric field is increased, after a critical value of the same, a peak at
$q\simeq 0$ is seen (see Fig.~$4(b)$). This is due to accumulation of the
charges near one side of the chain. At this value of the electric field, the
peak at  $q \simeq \pi$ starts to get diminished. For larger electric field,
the peak at $q\simeq 0$ becomes larger at the cost of the peak at $q \simeq
\pi$. In Fig.~$5(a)$ the variations of the magnitudes of the peaks at $q\simeq
0$ and $q \simeq \pi$ are shown. It clearly shows that the value of the
electric field at which the $q\simeq 0$ peak becomes significant is as same as
the value of the electric field where the first minimum of the charge gap
oscillation occurs (see Fig.~$3(a)$).  So it is clear that at this value of the
electric field charge accumulation at  one side of the chain begins to
dominate and the charge ordering tends to get destroyed. This point can  be
taken as a transition point, though on either sides of it the system remains
insulating. By observing the increase in height of the peak at $q\simeq 0$, one
can easily understand how the charge accumulation at one side of the chain
grows with electric field.\\

In Fig.~$3(b)$ we present the variation of the charge gap with the electric 
field for the other system, the 
$-0-0-U-$ superlattice at $2/3$ filling. At zero electric field, the
electrons try to accumulate at the sites with zero Hubbard interaction as it
minimizes the  energy. As a result, the system is charge ordered insulator
with a finite charge gap. A sharp peak of $C(q)$ at $q \simeq \frac{2 \pi}{3}$
is obtained. Here also the density dependent $2 k_F$ peak and the peak due to
the structural periodicity of the lattice occur at the same $q$ $( \simeq
\frac{2 \pi}{3})$. On application of the electric field, the charge gap
initially increases and then falls gradually. Subsequent oscillations are
observed as in the previous case of $-0-U-$ model.\\

The initial increase in the charge gap for small values of the
electric field can be understood in the following way. In a
particular cell of this $-0-0-U-$ superlattice at $2/3$ filling, the two sites
at the left (with $U=0$) are preferred by the electrons in the absence of
electric field; they tend to be doubly occupied keeping the other site
empty. In this situation a charge
density wave is formed in the chain. For nonzero electric field, there is
a positive gradient of the site potentials towards right (see Fig.~$1$). Then
the distribution of the site potentials in a unit cell
of the superlattice is in unison with the distribution of the correlation
parameter. This phenomenon reinforces the aforesaid charge ordering for  small
values of the electric field. So an increase in the value of the charge gap is 
observed. For larger electric field, however the spatial gradient of
the site potentials becomes so large, that a global shift of the charges
towards left is preferred in the chain and the charge ordering is
destroyed. Also for the $-0-U-$ model a trace of such feature is
observed for low electric field. There is a change in the slope of the charge
gap as a function of the applied voltage,
for a low value of the voltage (see Fig.~$3(a)$). Since in this case
the unit cell contains only two sites, the feature is only weakly showing
up. On the contrary, we have noted that for $-U-0-0-$ model and $-U-0-$ model,
this type of enhancement of charge ordering for small values of the electric
field is absent. In those cases, on application of the electric field the
charge gaps initially fall and then subsequent oscillations are observed.\\

\begin{figure}[h]
\resizebox{7.5cm}{!}
{\includegraphics*{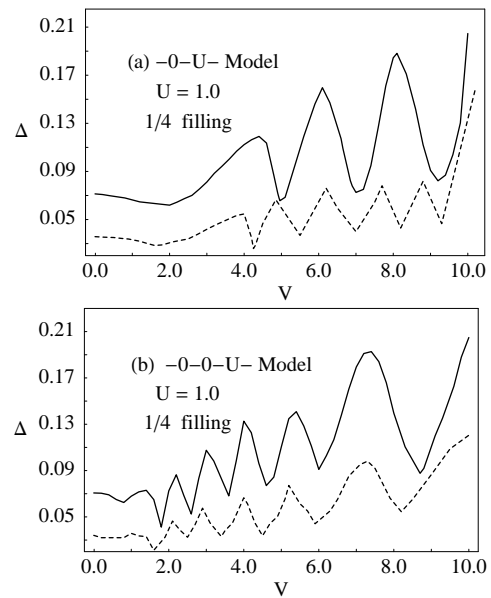}}
\caption{Variations of the charge gap $(\Delta)$ at the Fermi level for the
  quarter filled superlattice systems with the applied voltage $(V)$; (a) for 
$-0-U-$ superlattice and (b) for $-0-0-U-$ superlattice respectively. The solid
 line corresponds to $N=60$, while the dotted line shows the case of $N=120$.}
\end{figure}

\begin{figure}[h]
\resizebox{7.5cm}{!}
{\includegraphics*{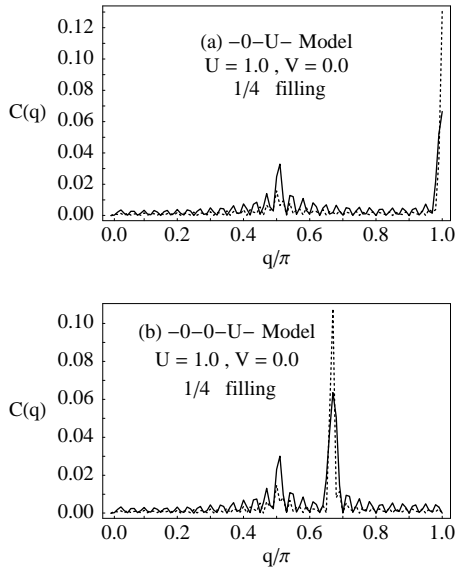}}
\caption{Charge density structure factor $C(q)$ for the quarter filled
  superlattice systems at zero electric field; (a) for $-0-U-$ superlattice
  and (b) for $-0-0-U-$ superlattice respectively. The solid line corresponds 
 to $N=60$, while the dotted line shows the case of $N=120$.}
\end{figure}

Figure~$5(b)$ 
shows the variation of the heights of the peaks of $C(q)$ with external bias
for the $-0-0-U-$ model; type of 
variation is quite similar to the previous case of $-0-U-$ model. The peak at
$q \simeq \frac{2\pi}{3}$ starts to fall and a peak at $q\simeq 0$ appears at a
critical value of the electric field. This critical value again matches with
the value of the electric field at which the first minimum of the charge gap
oscillation occurs in Fig.~$3(b)$. So in this case also it makes a transition
from a CDW phase to a phase where electrons pile up at one end.\\

We have shown the variation  of the charge gap for the $-0-0-U-$ superlattice 
at
 half-filling in Fig.~$3(c)$ for the sake of comparison with the other 
half-filled cases shown in  Fig.~$2$ and  Fig.~$3(a)$. Here at $V=0$ the gap 
diminishes with larger values of $N$ indicating a metallic behavior at the 
thermodynamic limit. An increase in $V$ drives the system ultimately to an 
insulating one.   

\begin{figure}[h]
\resizebox{7.5cm}{!}
{\includegraphics*{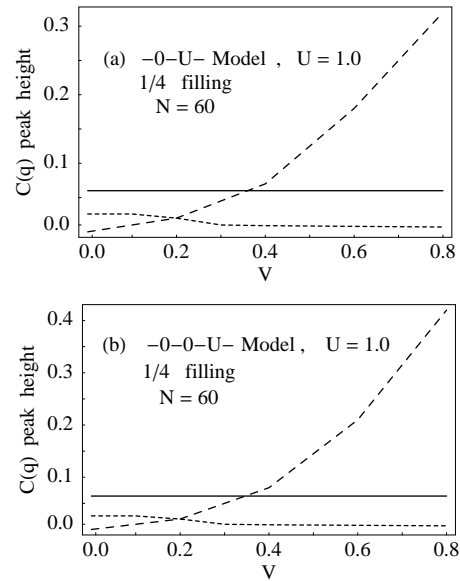}}
\caption{Variations of the height of peaks in $C(q)$ for the quarter
  filled superlattice systems with the applied voltage $(V)$; dashed line is
  for the peak at $q\simeq 0$, small dashed line is for the peak at
  $q \simeq \frac{\pi}{2}$ ($2k_F$) and solid
  line is for the peak due to structure (a) for $-0-U-$ superlattice and (b)
  for $-0-0-U-$ superlattice respectively.}
\end{figure}

We have also studied the quarter filled cases for our superlattice
models. Figures $6(a)$ and $6(b)$ show the charge gap oscillations in $-0-U-$
 and $-0-0-U-$ models respectively. Both the systems are metallic in
absence of the electric field; the charge gap $\Delta \sim \frac{1}{N}$, $N$ 
being
the system size. With the increase of electric field the charge gap oscillates
between a number of maxima and minima, and then increases monotonically,
indicating an insulating phase.\\

In Fig.~$7$ we plot the $C(q)$'s for both
types of the superlattices at quarter filling in absence of the electric
field. For $-0-U-$ model (Fig.~$7(a)$), we see a dominant peak at $q=\pi$,
which is due to the  structural periodicity of the lattice. Apart from this, a
$2k_{F}$ peak appears at $q \simeq \frac{\pi}{2}$.  Also few other
wiggles due to various possible short length scale density modulations are
observed. For $-0-0-U-$ model (Fig.~$7(b)$) also, the dominant peak is due to
the lattice structure and it occurs at $q=\frac{2 \pi}{3}$. Here also other
wiggles due to different possible density modulations are obtained along with
the $2k_{F}$ peak $(q \simeq \frac{\pi}{2})$.\\

It can be seen explicitly, in the limit $U=0$, that the position of the
$2k_{F}$ peak in finite chains is shifted from the expected position 
$q = 2 k_F = n \pi/(N+1)$  ($n$ being
the number of occupied single particle levels) by an amount $\sim
\frac{1}{N}$. As the system size $N$ increases, they approach the actual
$2k_{F}$ value. On the other hand, the peak heights fall as $\frac{1}{N}$. So
in the large $N$ limit (and also under periodic boundary condition) the 
$2k_{F}$ peaks are not detectable.\\

 As the electric
field is turned on and increased, a  peak at $q \simeq 0$ appears for both the
models at quarter filling. The peak due to the structural periodicity and the 
$2k_{F}$ peak (also
the wiggles in $C(q)$) fall off; rate of fall is much slower for the structural
periodicity peak. These show the gradual accumulation of the charges near one
side of the chain. In Fig.~$8$ variations  of the heights of the peaks in
$C(q)$'s with the external bias is shown. It clearly  shows how charge
accumulation at one side of the chain increases with electric field.\\

\section{Conclusion}

In this work, we have studied the one dimensional Hubbard superlattices with
different types of unit cells in the presence of electric field. Single orbital
nearest neighbor tight-binding model has been used. We have maintained the 
fixed
system sizes ($N=60$ and $120$), for the purpose of comparison of these models.
 The 
charge gap and the CDW structure factor of the systems are studied under 
Hartree-Fock approximation in the
presence of electric field. To check the reliability of the
Hartree-Fock approximation results, we compared our results for the
homogeneous Hubbard model with a previous DMRG
calculation\cite{Pati2} and found reasonable qualitative agreement. 
Oscillations in charge
gap obtained in the superlattice systems are  rather similar to those observed
in the homogeneous Hubbard chains, showing signature of the finite sizes.
 Variations of the heights of different peaks of $C(q)$ with the
applied electric field give an idea about the distribution of electrons on the
lattice. We found that the $2k_F$ peaks of $C(q)$ arise for finite sized
systems only. On application of the electric field a peak at $q \simeq 0$ 
appears at a critical
value of the field and then increases in height; other peaks fall
gradually indicating suppression of the ordering due to the Hubbard
correlation on the superlattice. The rate of fall of the peak corresponding to
the structural periodicity is much slower than the $2k_F$ peak. At the critical
value of the electric field where the peak at $q \simeq 0$ appears in $C(q)$,
the first minimum of the charge gap is observed for the systems which were
initially charge ordered. The present mean field approximation is capable of
detecting the variations of the charge structure with the electric field. One
can also use other methods to study these features. We have presented here
only two types of superlattices with few different fillings. For other
superlattice structures and for other fillings also the effect of the electric
field may be explored. The superlattices at finite temperatures may reveal
some interesting features on application of the electric field.

\begin{acknowledgments}
Authors sincerely acknowledge useful discussions with S. Sil.
\end{acknowledgments}

\end{document}